\documentclass[11pt]{article}
\RequirePackage[OT1]{fontenc}
\RequirePackage{amsthm,amsmath}
\RequirePackage{natbib}

\usepackage{amsmath}
\usepackage{graphicx,psfrag,epsf,subfig}
\usepackage{enumerate}
\usepackage{natbib}
\usepackage[graphicx]{realboxes}
\usepackage{url}

\usepackage{multicol, booktabs, subfig, bbm, color, xcolor}
\usepackage{amssymb}

\usepackage[margin=1in]{geometry}
\usepackage{setspace}

\usepackage{titlesec}

\titleformat*{\section}{\normalsize\bfseries}
\titleformat*{\subsection}{\normalsize\it}
\usepackage[affil-it]{authblk}

\graphicspath{{figs/}}

\title{\large\bfseries Kernel Machine and Distributed Lag Models for Assessing Windows of Susceptibility to Environmental Mixtures in Children's Health Studies}
\date{}

\author[1]{Ander Wilson\footnote{ander.wilson@colostate.edu}}
\author[2]{Hsiao-Hsien Leon Hsu}
\author[2]{Yueh-Hsiu Mathilda Chiu}
\author[2]{Robert O. Wright}
\author[2]{Rosalind J. Wright}
\author[3]{Brent A. Coull}
\affil[1]{Colorado State University}
\affil[2]{Icahn School of Medicine at Mount Sinai}
\affil[3]{Harvard T. H. Chan School of Public Health}

\begin{document}

\maketitle

\begin{abstract}
Exposures to environmental chemicals during gestation can alter health status later in life. Most studies of maternal exposure to chemicals during pregnancy have focused on a single chemical exposure observed at high temporal resolution. Recent research has turned to focus on exposure to mixtures of multiple chemicals, generally observed at a single time point. We consider statistical methods for analyzing data on chemical mixtures that are observed at a high temporal resolution.  As motivation, we analyze the association between exposure to four ambient air pollutants observed weekly throughout gestation and birth weight in a Boston-area prospective birth cohort. To explore patterns in the data, we first apply methods for analyzing data on (1)  a single chemical observed at high temporal resolution, and (2) a mixture measured at a single point in time. We highlight the shortcomings of these approaches for temporally-resolved data on exposure to chemical mixtures.  Second, we propose a novel method, a Bayesian kernel machine regression distributed lag model (BKMR-DLM),   that simultaneously accounts for nonlinear associations and interactions among time-varying measures of exposure to mixtures.  BKMR-DLM uses a functional weight for each exposure that parameterizes the window of susceptibility corresponding to that exposure within a kernel machine framework that captures non-linear and interaction effects of the multivariate exposure on the outcome. In a simulation study, we show that the proposed method can better estimate the exposure-response function and, in high signal settings, can identify critical windows in time during which exposure has an increased association with the outcome. Applying the proposed method to the Boston birth cohort data, we find evidence of a negative association between organic carbon and birth weight and that nitrate modifies the organic carbon, elemental carbon, and sulfate exposure-response functions.
\end{abstract}

\section{Introduction}\label{sec:intro}

Humans are inevitably exposed to a complex mixture of chemicals and other pollutants throughout the life course beginning with conception \citep{Woodruff2011,Wright2017}. Epidemiological evidence about the toxicity of environmental chemicals has traditionally come from studies of a single exposure observed during a single time window,  such as averaged over a pre-specified time period. The one-chemical-at-a-time and one-exposure-window-at-a-time approaches can result in misleading estimates by failing to distinguish between the effects of multiple highly correlated chemical exposures \citep{Braun2016} or by incorrectly identifying the time window during which someone is vulnerable to a chemical exposure \citep{Wilson2017}, respectively. It is, therefore, critical that statistical methods be able to handle exposure data for mixtures of multiple chemicals for which temporally-resolved measurements reflect changing exposure levels throughout the life course.

In the study of the risks associated with maternal exposures to air pollution during pregnancy and children's health, there is particular interest in exposure timing. Two primary goals of these analyses are to identify windows of susceptibility, which are periods during which an exposure can influence a future birth or child's health outcome, and to estimate the exposure-response relationship. Popular statistical methods typically use time-resolved measures of exposure, such as average exposure for each week of pregnancy, to identify windows of susceptibility.  Recent research has identified windows of susceptibility and estimated the exposure-response relationship between prenatal air pollution exposure and lower birth weight, increased risk of preterm birth, and decreased childhood respiratory health, among other outcomes \citep{Chang2012,Warren2013,Hsu2015,Bose2017a,Lee2017}. 

Partly due to a dearth of available methods, all of these studies estimate the association between time-resolved measurements of a single environmental chemical and a health outcome. There exists a gap with respect to methods for analyzing the association between health and time-resolved measurements of a chemical mixture. Assessing the relationship between  time-resolved measures of an environmental mixture and a health outcome is complicated by several factors. These include: 1) high correlation among multiple chemical exposures at each time point; 2) high temporal correlation among repeated measures of a given exposure; 3) potential nonlinear associations between any given chemical exposure and the health endpoint; and 4) potential interactions in the health effects of multiple chemicals, either within or across exposure times. Statistical approaches have been proposed to address each of these challenges individually. However, no approach that fully captures the nonlinear and non-additive effects of temporally-resolved exposure measurements on health currently exists.

\subsection{Relevant statistical literature}

The proposed work in this paper relates to two active areas of statistical methods development. The first area is methods for lagged exposures. These methods relate a scalar health outcome to time-resolved measures of an exposure observed prior to assessment of the health outcome. The second area is statistical methods to estimate the health effects of chemical mixtures. To date, very little research has focused on statistical methods for mixtures of time-resolved exposures.

Recent statistical methods developed to estimate the association between maternal exposures during pregnancy and a birth outcome have focused on distributed lag models (DLMs). In general, a DLM regresses an outcome observed at a single time point on exposures observed on an evenly spaced grid over a preceding period of time. The effect of exposure on the outcome is typically constrained to vary smoothly in time. This smoothness constraint regularizes the model in the presence of high temporal autocorrelation in the exposure measures. Recent approaches have formulated the smooth distributed lag effect using splines, Gaussian processes and principal components \citep{Warren2012,Chang2015,Warren2016,Wilson2017a,Gasparrini2017,Warren2019}. Distributed lag nonlinear models (DLNMs) have been proposed to extend the DLM to nonlinear associations \citep{Gasparrini2010,Gasparrini2011,Gasparrini2017,Mork2020}. Several DLM methods have been proposed for two time-resolved predictors. This includes both additive models \citep{Warren2013} and models with interactions \citep{Chen2018}. These DLM methods have been developed for at most two pollutants. 

For mixtures of more than two chemicals, numerous statistical methods have been proposed to estimate the association between exposure observed at a single time point and a health outcome. Proposed methods include Bayesian nonparametric shrinkage and selection priors \citep{Herring2010}, clustering approaches \citep{Molitor2010,Zanobetti2014,Pearce2014}, exposure index methods \citep{Carrico2014,Park2014,Keil2020}, and exposure-response surface methodology \citep{Bobb2015}. Of particular relevance to the current paper is Bayesian kernel machine regression (BKMR), which estimates a flexible, high-dimensional exposure-response surface \citep{Bobb2015}. For recent reviews of statistical methods for chemicals mixtures see \cite{Taylor2016b,Davalos2017,Hamra2018} and \cite{Gibson2019}.  All of these methods estimate the relationship between a health outcome and exposure to a chemical mixture measured at a single point in time, or averaged over a single pre-specified exposure window. 

For studying the effect of exposure to mixtures of three or more chemicals, methods to handle exposures measured at multiple time points have focused on a small number of discrete time points. \cite{Liu2018a} developed lagged kernel machine regression for mixtures observed at multiple points. The approach is appropriate for exposures observed at a small number of times that are common to all individuals in the study (e.g. blood biomarkers measured once per trimester). However, it does not scale to finer temporal-resolution exposure data such as weekly exposure throughout pregnancy. Another limitation with respect to the current setting is the approach only estimates interactions among exposures at the same time point. It cannot estimate interactions across time. \cite{Bello2017} proposed a lagged weighted quantile sum regression model. The approach regresses the time-resolved exposures on the outcome to estimate the association but does not account for nonlinearities or interactions. No methods have been proposed that fully integrate methods for mixtures with those for finely-resolved temporal measures of exposure.

\subsection{Our Contribution}

We propose a framework for estimating the relationship between time-resolved measures of an environmental mixture and a health outcome. We refer our new framework as a Bayesian kernel machine regression distributed lag model (BKMR-DLM). BKMR-DLM uses time-weighted exposures \citep{Wilson2017a} to reduce the dimension of the time-resolved exposure data and to identify windows of susceptibility. The potentially nonlinear and non-additive association between these time-weighted exposures and a health outcome is modeled using kernel machine regression. The kernel machine is a penalized estimator that can reduce the effect of multicollinearity between the multiple exposures. Taken together, the model represents a new form of structured multiple index model (MIM) \citep{Xia2008a} that imposes structure on the weight functions that form linear combinations of covariates, which then serve as inputs into a nonparametric function.  To handle the larger parameter space required to account for exposure timing and the parameter constraints necessary for MIMs, we propose a new MCMC algorithm for the BKMR framework. Hence, BKMR-DLM integrates modern methods for time-resolved measures of exposures and mixtures in a coherent Bayesian framework. To our knowledge, this is the first approach to simultaneously address nonlinearities, interactions, and exposure timing in a single cohesive model.

Section~\ref{s:data} presents the motivating data and the results of preliminary analyses of these data using standard methods. These exploratory analyses introduce notation and background, illustrate some general patterns in the data, and illustrate limitations of existing methodology. Section~\ref{s:kmrwithtime} presents the proposed methods. Section~\ref{s:sim} compares the proposed approach to established methods in a simulation study. We show that BLMR-DLM can estimate windows of susceptibility to components of a mixture and that BKMR-DLM can more accurately estimate exposure-response functions than established DLM or mixtures methods applied to exposure summaries. In Section~\ref{s:da} we apply BKMR-DLM to data from a birth cohort study conducted in Eastern Massachusetts, USA, to estimate the association between weekly maternal exposure to four pollutants and  birth weight for gestational age $z$-score (BWGAz). We conclude with a discussion in Section~\ref{s:discussion}.


\section{Data, Notation, and Preliminary Analyses}\label{s:data}

\subsection{Data}
The Asthma Coalition on Community, Environment, and Social Stress (ACCESS)  cohort \citep{Wright2008}  is a prospective, longitudinal study designed to examine the effects of psychosocial stressors and chemical stressors (e.g., air pollution and other environmental influences) on children's birth and health outcomes. ACCESS includes 955 mother-child dyads recruited between August 2002 and January 2007 who continued active follow-up after birth in the Boston, MA area. Procedures were approved by the human studies committees at Brigham and Women’s Hospital and Boston Medical Center.

Previous analyses of the ACCESS cohort identified an association between increased maternal exposure to fine particulate matter (PM$_{2.5}$) averaged over pregnancy and decreased birth weight for gestational age $z$-score (BWGAz), particularly among boys born to obese mothers \citep{Lakshmanan2015}. In this paper we consider the association between weekly levels of exposure to four components of particulate matter--elemental carbon (EC), organic carbon (OC), nitrate, and sulfate--and BWGAz among the same population of boys with obese mothers. We include as covariates maternal age at enrollment, an indicator of maternal education at high school level or above,  maternal pre-pregnancy body mass index, indicators of black and Hispanic race/ethnicity, parity, and an indicator of season of birth. 

Maternal exposures of EC, OC, nitrate, and sulfate were  estimated with a hybrid land use regression model that incorporates satellite-derived aerosol optical depth measures and a chemical-transport model GEOS-Chem \citep{Di2016}. Each mother was assigned an average exposure level for each pollutant for each week of pregnancy based on the predicted value at her address of residence. We limit our analysis to full-term infants (born at $\ge 37$ weeks gestation) and their exposure during the first 37 weeks of pregnancy. A total of 109 children had complete exposure, outcome, and covariate data. For completeness, we show results for the full cohort in the supplemental material.

\subsection{Objectives and notation}

Interest focuses on estimating the association between time-resolved measures of a mixture of $M$ pollutants and a scalar outcome $Y$, while controlling for a $p$-vector of baseline covariates $\mathbf{Z}$. We denote the exposures at time $t$ as $X_{1}(t),\dots,X_{M}(t)$ for $t\in\mathcal{T}$. In our analysis $\mathcal{T}=\{1,\dots,37\}$ is the first 37 weeks of gestation.  We assume these quantities are observed for a sample of size $n$ with subject indexed by $i$. 

There are two primary objectives: 1) identify windows of susceptibility during which exposure to a chemical is associated with a future health outcome and 2) estimate the exposure-response relationship while allowing for a nonlinear and non-additive relationship between the multiple  exposures and the outcome. 

\subsection{Preliminary analysis with additive DLM}\label{sub:dlm}

The Gaussian discrete-time DLM for a single exposure is 
\begin{equation}
Y_i=\alpha + \sum_{t=1}^T X_{it}\delta(t)+\mathbf{Z}_i^T\boldsymbol{\gamma}+\epsilon_i,
\label{eq:DLM}
\end{equation}
where $\delta(t)$ parameterizes the association between exposure at time $t$ and the outcome. In \eqref{eq:DLM}, $\alpha$ is the intercept and $\boldsymbol{\gamma}$ is a $p$-vector of unknown regression coefficients. The $\epsilon_i$'s are assumed to be iid $\text{N}(0,\sigma^2)$. The scalar outcome $Y_i$ is assumed to be observed post exposure.

Constrained DLMs impose smoothness on the distributed lag function $\delta(t)$. The smoothness constraint can be imposed by modeling $\delta(t)$ using splines, Bayesian priors, Gaussian processes, or other penalization approaches \citep{Zanobetti2000,Peng2009,Heaton2012,Chen2017}. In this paper,  we use natural splines to impose smoothness on $\delta(t)$ and select the degrees of freedom as the value that minimizes the Akaike information criterion (AIC).

\begin{figure}
    \centering
    \includegraphics[width=\textwidth]{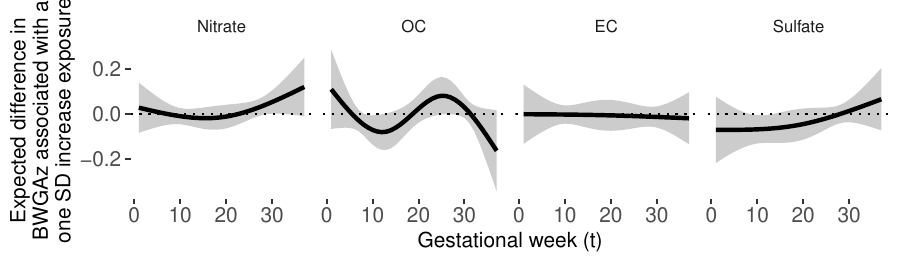}
    \caption{Estimated distributed lag function between exposures and birth weight for gestational age $z$-score in ACCESS using the additive DLMs. This is $\hat\delta(t)$ estimated from \eqref{eq:DLM}. The function represents the estimated expected change in BWGAz per one standard deviation change in exposure ($y$-axis) as a function of gestational week ($x$-axis). }
    \label{fig:dlmwindows}
\end{figure}

For multiple exposures, an additive DLM is 
\begin{equation}
Y_i=\alpha + \sum_{m=1}^M\sum_{t=1}^T X_{imt}\delta_{m}(t)+\mathbf{Z}_i^T\boldsymbol{\gamma}+\epsilon_i. 
\label{eq:additiveDLM}
\end{equation}
Figure~\ref{fig:dlmwindows} shows results from the additive DLM analysis of the ACCESS data.  We found suggestive evidence of susceptibility windows in weeks 29-33 for nitrate, in weeks 9-12 for OC, and in weeks 9-13 in sulfate. Susceptibility windows are defined as times when the pointwise 95\% confidence interval does not contain zero. There was moderate evidence of a cumulative effect, representing the change in birth weight associated with a one unit increase in exposure at every time point, for sulfate  ($p$-value$=0.07$) but not for any other pollutant. 

An additive DLM is appealing because it is easy to visualize and interpret, but does not allow for nonlinear associations or interactions among exposures. An additive DLNM allows for nonlinear associations but not interactions. Moreover, the nonlinear extension makes identification of critical windows more complicated, as the definition of a window depends on the level of the exposure.

\subsection{Preliminary mixtures analysis with BKMR}\label{sub:kmr}

BKMR is a popular approach to estimate the association between multiple scalar exposures and a health outcome. BKMR allows for nonlinear associations and interactions among exposures. 

For $M$ scalar exposures $\mathbf{E}_i=(E_{i1},\dots,E_{iM})^T$, a BKMR model takes the form
\begin{equation}
Y_i = h(E_{i1},\dots,E_{iM}) + \mathbf{Z}_i^T\boldsymbol{\gamma} + \epsilon_i. 
\label{eq:bkmrsingle}
\end{equation}
The function $h(\cdot)$ is a potentially nonlinear and non-additive exposure-response function. 

BKMR assumes that the exposure-response function $h:\mathbb{R}^M\rightarrow\mathbb{R}$ resides in the functional space $\mathcal{H}_K$ that is uniquely defined by the positive semidefinite reproducing kernel $K:\mathbb{R}^M\times\mathbb{R}^M\rightarrow\mathbb{R}$. The function $h(\cdot)$ can be  represented with a positive-definite kernel function $K(\cdot,\cdot)$ and coefficients $\{\alpha_i\}_{i=1}^n$ as $h(\mathbf{E})=\sum_{i=1}^n K(\mathbf{E},\mathbf{E}_i)\alpha_i$.  According to Mercer's Theorem \citep{Cristianini2000}, the kernel $K(\cdot,\cdot)$ implicitly specifies a  basis expansion. For example, the Gaussian kernel corresponds to the set of Gaussian radial basis functions. 

Using the kernel representation  of $h(\cdot)$, \cite{Liu2007} showed that  the regression model in \eqref{eq:bkmrsingle} is equivalent to the hierarchical model 
\begin{eqnarray}
\label{eq:hiermodel}
Y_i &\sim&\text{N}(h_i+\mathbf{Z}_i^T\boldsymbol{\gamma}, \sigma^2)\\ 
\mathbf{h}=\left(h_1,\dots,h_n\right)^T &\sim&\text{N}(\mathbf{0},\sigma^2\tau^2\mathbf{K}),\nonumber
\end{eqnarray}
where $\mathbf{K}$ is an $n\times n$ matrix with $i,j$ element $K_{ij}=K(\mathbf{E}_i,\mathbf{E}_j)$. For a Gaussian kernel $K_{ij}=\exp[-\sum_{m=1}^M \rho_m(E_{im}-E_{jm})^2]$. 

In our preliminary mixtures analysis with BKMR, we take $E_{im}$ to be the average exposure level for pollutant $m$ over the first 37 weeks of pregnancy. All time points are given equal weight in the model. Hence, BKMR does not account for exposure timing. We fit the model with a Gaussian kernel as implemented in the R package {\tt bkmr} \citep{Bobb2017,Bobb2018} using the default hyperparameters for all prior distributions. 

\begin{figure}[h!]
    \centering
    \includegraphics{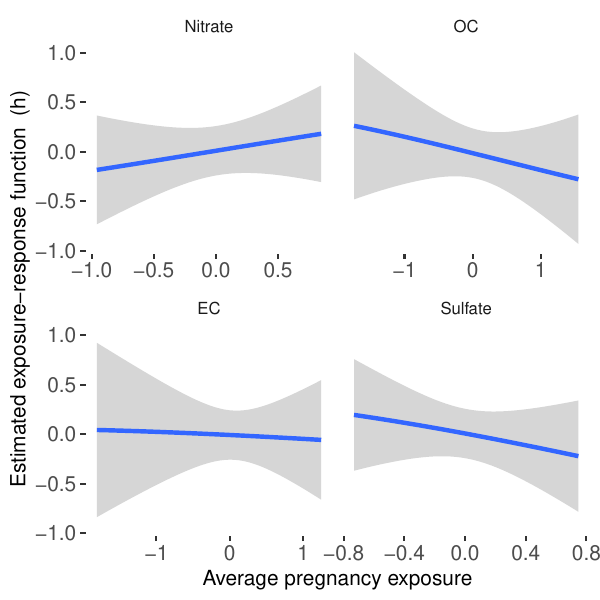}
    \caption{Estimated exposure-response function with BKMR. The exposure-response function is shown for each pollutant at the median value of all other pollutants. The $x$-axis is mean exposure over the first 37 weeks of gestation using exposure values standardized to have mean zero and variance one.}
    \label{fig:bkmrmain}
\end{figure}

Figure~\ref{fig:bkmrmain} shows results from applying BKMR to the ACCESS data using average exposures of each pollutant over the first 37 weeks of gestation. Each panel shows the estimated exposure-response surface between one pollutant and BWGAZ while holding the other three pollutants fixed at their median value. Hence, the figure shows negative associations between OC, EC, sulfate and BWGAZ but a positive association with nitrate. All of the intervals contain the null association. Supplemental Figure S1 shows that there is no evidence of interaction between pairs of pollutants.

\section{BKMR-DLM  Model Specification} \label{s:kmrwithtime}

The BKMR approach outlined in Section~\ref{sub:kmr} centers on specification of a kernel function that has as inputs scalar measures of the exposures. Replacing the scalar exposures $E_{i1},\dots,E_{iM}$ with the time-resolved exposures $\{X_{i1}(1) ,\dots,X_{i1}(T),X_{i2}(1),\dots,X_{iM}(T)\}$ poses two major challenges. First, there is a dimensionality issue. This comes from replacing the vector of $M$ scalar exposures with the $M\times T$ exposure measures. Second, the model does not impose any structure with respect to how the effect of an exposure varies over time. This structure is needed both to conform with our biological understanding that the exposure effect in proximal weeks should be similar and to add stability in the presence of temporal autocorrelation. 

The central concept of the proposed approach is to include the time-resolved exposures into the kernel function and add structure with a weight function. The weight function up- and down-weights time periods so that they have a larger or smaller influence on the exposure-response function. The weight function can be interpreted in a fashion similar to that of a traditional DLM. Up-weighted time periods define the window of susceptibility. 

The weight function for exposure $m$ is $w_m(t)$ and is defined over the domain $\mathcal{T}$. The weighted exposure is  $E_{im}=\int_\mathcal{T} X_{im}(t)w_m(t)dt$, where $X_i(t)$ is a functional  representation of the exposure. We replace the scalar predictor  in the kernel function with the weighted exposure. The Gaussian kernel is then
\begin{equation}
K_{ij}=\exp\left[-\sum_{m=1}^M \rho_m\left\{\int_\mathcal{T} X_{im}(t)w_m(t)dt-\int_\mathcal{T} X_{jm}(t)w_m(t)dt\right\}^2\right].
\label{eq:kernelfunc}
\end{equation}

In addition to the Gaussian kernel we also implement a polynomial kernel. This provides a more parsimonious representation of the exposure-response function. The polynomial kernel is less flexible but potentially more efficient when the simpler structure holds \citep{Liu2007}. The polynomial kernel of order $d$ is 
\begin{eqnarray}
K_{ij}&=&\left(1+\sum_{m=1}^M \rho_m E_{im}E_{jm}\right)^d \label{eq:kernelscalarquad}\\
&=&\left[1+\sum_{m=1}^M \rho_m \left\{\int_\mathcal{T} X_{im}(t)w_m(t)dt\right\}\left\{\int_\mathcal{T} X_{jm}(t)w_m(t)dt\right\}\right]^d. \nonumber
\end{eqnarray}

For either kernel, the sign of the weights $w_m(t)$ is not identifiable as it could be replaced with $-w_m(t)$ and result in the same likelihood. The magnitude of $\rho_m$ and $w_m(t)$ are not individually identifiable as they could be scaled by the same factor and result in equal likelihood. We impose the identifiability constraints $\int_\mathcal{T} w_m(t)dt>0$ to identify the sign of $w_m(t)$ and $\int_\mathcal{T} \left[w_m(t)\right]^2dt=1$ to identify the magnitude of $w_m(t)$ and $\rho_m$. Under these two constraints, both $\rho_m$ and $w_m(t)$ are identifiable.

\subsection{Relation to DLM and BKMR}

The weight function approach used in BKMR-DLM is closely  related to DLMs and functional regression methods. In order to see the connection, we write a DLM for a single exposure as
\begin{equation}
Y_i=\alpha + \beta\int_\mathcal{T} X_i(t)w(t)dt+\mathbf{Z}_i^T\boldsymbol{\gamma}+\epsilon_i.
\label{eq:bdlim}
\end{equation} 
The model can be viewed as a linear regression model using the weighted exposure $E_i=\int_\mathcal{T} X_i(t)w(t)dt$ as a scalar covariate: $Y_i=\alpha + \beta E_i +\mathbf{Z}_i^T\boldsymbol{\gamma}+\epsilon_i$. This is equivalent to the functional DLM $Y_i=\alpha + \int_\mathcal{T} X_i(t)\delta(t) dt+\mathbf{Z}_i^T\boldsymbol{\gamma}+\epsilon_i$ with functional predictor $\delta(t)=\beta w(t)$. With a linear kernel and a single predictor the BKMR-DLM approach reduces to this basic DLM.

When $w(t)$ is constant over time, BKMR-DLM is equivalent to a BKMR model that uses  pregnancy-averaged exposure as a predictor. When $w(t)$ varies over time,  BKMR-DLM up- or down-weights exposures during certain time periods. The up-weighted time periods represent windows of susceptibility to exposure.

\subsection{Parameterization of the weight function}

For each exposure, we parameterize both $X_{im}(t)$ and $w_m(t)$ using a basis function representation \citep{Morris2015a}. We assume $X_{im}(t)=\sum_{l=1}^{L_m}\xi_{iml}\psi_{ml}(t)$ and $w_m(t)=\sum_{l=1}^{L_m}\theta_{ml}\psi_{ml}(t)$.  Both $X_{im}(t)$ and $w_m(t)$ are parameterized  with the same orthonormal basis $\{\psi_{ml}(t)\}_{l=1}^{L_m}$. The regression coefficients for $X_{im}(t)$ and $w_m(t)$ are $\{\xi_{iml}\}_{l=1}^{L_m}$ and $\{\theta_{ml}\}_{l=1}^{L_m}$, respectively. The $m^\text{th}$ weighted exposure for individual $i$ can then be rewritten as $E_{im}=\boldsymbol{\xi}_{im}^T\boldsymbol{\theta}_m$ where $\boldsymbol{\theta}_m=\left[\theta_{m1},\dots,\theta_{mL_m}\right]^T$ and $\boldsymbol{\xi}_{im}=\left[\xi_{im1},\dots,\xi_{imL_{m}}\right]^T$. We estimate $\boldsymbol{\xi}_{im}$ using ordinary least squares which gives $E_{im}=\mathbf{X}^T_{im}\boldsymbol{\Psi}_m\boldsymbol{\theta}_m$, where $\mathbf{X}^T_{im}$ is the row-vector of observed exposures for chemical $m$ and person $i$ and $\boldsymbol{\Psi}_m$ is the design matrix of orthonormal basis functions.

Using an orthonormal basis, the constraint $\int_\mathcal{T}  [w_m(t)]^2dt=1$ is satisfied if and only if $\|\boldsymbol{\theta}_m\|=1$. The constraint $\int_\mathcal{T} w_m(t)dt\ge0$ is satisfied for a set of observed times if and only if $\mathbf{1}^T\boldsymbol{\Psi}_m\boldsymbol{\theta}_m\ge0$, where $\mathbf{1}$ is a vector of ones. As such, the constraints on $w_m(t)$ are now constraints on $\boldsymbol{\theta}_m$. The constrained parameter space is half of a unit $K_m$-ball on one side of a hyperplane defined by $\mathbf{1}^T\boldsymbol{\Psi}_m\boldsymbol{\theta}_m=0$.

Using the weighted exposures as inputs, the Gaussian kernel function in \eqref{eq:kernelfunc} is 
\begin{equation}
K_{ij} =  \exp\left[-\sum_{m=1}^M\rho_m\left\{(\mathbf{X}_{im}-\mathbf{X}_{jm})^T\boldsymbol{\Psi}_m\boldsymbol{\theta}_m\right\}^2\right],
\label{eq:K1}
\end{equation}
and the polynomial kernel in \eqref{eq:kernelscalarquad} is 
\begin{equation}
K_{ij}  = \left[1+\sum_{m=1}^M\rho_m(\mathbf{X}_{im}\boldsymbol{\Psi}_m\boldsymbol{\theta}_m)(\mathbf{X}_{jm}\boldsymbol{\Psi}_m\boldsymbol{\theta}_m)\right]^d.
\label{eq:K2}
\end{equation} 
The parameters $\rho_m$ and $\boldsymbol{\theta}_m$ represent the importance and the timing of exposure $m$, respectively.  To ease computation,  we reparameterize the model in terms of  $\boldsymbol{\theta}^*_m=\boldsymbol{\theta}_m\rho_m^{1/2}$. The Gaussian kernel in \eqref{eq:kernelfunc} is then \begin{equation}
K_{ij}  = \exp\left[-\sum_{m=1}^M\left\{(\mathbf{X}_{im}-\mathbf{X}_{jm})^T\boldsymbol{\Psi}_m\boldsymbol{\theta}^*_m\right\}^2\right].
\label{eq:newkernel}
\end{equation}
The polynomial kernel in \eqref{eq:kernelscalarquad} can be written as
\begin{equation}
K_{ij}  = \left[1+\sum_{m=1}^M(\mathbf{X}_{im}\boldsymbol{\Psi}_m\boldsymbol{\theta}^*_m)(\mathbf{X}_{jm}\boldsymbol{\Psi}_m\boldsymbol{\theta}^*_m)\right]^d.
\label{eq:newkernelquad}
\end{equation}

Using this formulation, both $\rho_m$ and $\boldsymbol{\theta}_m$ are identified by $\boldsymbol{\theta}_m^*$ as $\rho_m^{1/2}=||\boldsymbol{\theta}_m^*||$ and $\boldsymbol{\theta}_m=\rho_m^{-1/2}\boldsymbol{\theta}_m^*\text{sign}\{\int_\mathcal{T} w_m(t) dt \}$. Hence, we can estimate the full model parameterized in terms of $\boldsymbol{\theta}^*$ and then partition the posterior sample of $\boldsymbol{\theta}^*_m$ into $\rho_m$ and $\boldsymbol{\theta}_m$,  where $\boldsymbol{\theta}_m$ describes the weight function $w_m(t)$.

To induce smoothness in the weight function we use the eigenfunctions of the covariance matrix of smoothed exposures. We pre-smooth each exposure with a parsimonious natural spline bases and then use the eigenfunctions of the covariance matrix of the smoothed exposures in the model as specified above. Ideally a rich basis expansion would be used. This will increase the resolution with which a critical window can be identified. In practice, we have found that the method performs best with a parsimonious basis. We  used natural splines with 4 degrees of freedom. Increasing the degrees of freedom results in increased variance of the estimated weight functions and is to only be supported in unrealistically high signal-to-noise settings.

\subsection{Prior specification and posterior computation}

We use as a prior for $\boldsymbol{\theta}_m$, $m=1,\dots,M$, a  uniform distribution over its parameter space, which can be written as $p(\boldsymbol{\theta}_m)\propto\exp\left(-\boldsymbol{\theta}_m^T\boldsymbol{\theta}_m/2\right)\mathbbm{1}(\boldsymbol{\theta}_m^T\boldsymbol{\theta}_m=1)\mathbbm{1}(\mathbf{1}_{L_m}^T\boldsymbol{\theta}_m>0)$, where $\mathbbm{1}(\cdot)$ is an indicator function. We then let $\rho_m/\kappa_m\sim\chi^2_1$ for fixed value $\kappa_m$. It follows that $\boldsymbol{\theta}_m^*=\rho_m^{1/2}\boldsymbol{\theta}_m\sim\text{N}(0,\nu_m\kappa_m\mathbf{I}_{L_m})$, with $\nu_m\sim\chi^2_1$. We complete the prior specification by assuming a flat prior on $\boldsymbol{\gamma}$, $\sigma^{-2}\sim\text{gamma}(a_1,b_1)$, and $\log(\tau^2)\sim\text{N}(0,b)$.

\cite{Bobb2015} updated each of the $M$ parameters in the kernel function independently with Metropolis-Hastings. This approach is unappealing for our model as we have $\sum_{m=1}^M L_m$ parameters in the kernel function and potentially high correlation among parameters due to the temporal correlation in the exposures. We first integrate $\mathbf{h}$, $\boldsymbol\gamma$ and $\sigma^{-2}$ out of \eqref{eq:hiermodel} to obtain the marginalized posterior $p(\boldsymbol{\theta}_1^*,\dots,\boldsymbol{\theta}_M^*,\tau^2,\nu_1,\dots,\nu_M|\mathbf{Y})$. Then, our MCMC algorithm iteratively samples each $\boldsymbol{\theta}_m^*$ as a block using an elliptical slice sampler \citep{Murray2009} and the kernel of the marginalized posterior. Then we sample $\tau^{-2}$ using random walk Metropolis-Hastings using the same marginalized posterior. Finally, we use a Gibbs sampler to simulate $\sigma^{-2}$, $\boldsymbol{\gamma}$, and $\nu_1,\dots,\nu_M$ from their respective full conditionals.  Supplemental Section B provides additional details, full conditionals, and the full algorithm.

\subsection{Posterior inference for $w(t)$}

Windows of susceptibility during which there is an increased association between exposure and outcome are identified using the estimated weight function. Let  $\boldsymbol{\theta}^{*(r)}_m$ for $r=1,\dots,R$ be the posterior sample of size $R$. We can identify $\boldsymbol{\theta}^{(r)}_m=\boldsymbol{\theta}^{*(r)}_m\|\boldsymbol{\theta}^{*(r)}_m\|^{-1}\text{sign}(\mathbf{1}_{L_m}^T\boldsymbol{\theta}^{*(r)}_m)$ and $w_m^{(r)}(t)=\sum_{l=1}^{L_m} \theta^{(r)}_{ml}\psi_{ml}(t)$. The credible interval provides valid pointwise posterior inference for $w(t)$ and can be used to identify windows of susceptibility. However, the posterior mean does not satisfy the constraint $\int_\mathcal{T}[w(t)]^2dt=1$. We use the point estimate projected onto the parameter space of $\boldsymbol{\theta}_m$: $\widehat{w}_m(t)=\mathbf{X}^{*T}_{im}\widehat{\boldsymbol{\theta}}_m$ with $\widehat{\boldsymbol{\theta}}_m=\bar{\boldsymbol{\theta}}_m\|\bar{\boldsymbol{\theta}}_m\|^{-2}$ and $\bar{\boldsymbol{\theta}}_m$ is the posterior mean. The resulting estimator, equivalent to the  Bayes estimate with respect to the loss function $L(\boldsymbol{\theta}_m,\widehat{\boldsymbol{\theta}}_m) = [(\boldsymbol{\theta}_m-\widehat{\boldsymbol{\theta}}_m)^T(\boldsymbol{\theta}_m-\widehat{\boldsymbol{\theta}}_m)]/\mathbbm{1}\{\|\widehat{\boldsymbol{\theta}}_m\|=1\}$, is a central estimate in the parameter space of $\boldsymbol{\theta}_m$.

Estimates of $h(\cdot)$ for the observed exposure levels can be obtained by sampling from the conditional distribution of $\mathbf{h}$ from \eqref{eq:hiermodel} \citep{Bobb2015}. Full details are in Supplemental Section B.


\section{Simulation Study}\label{s:sim}

\subsection{Simulation design}

We evaluate the effectiveness of BKMR-DLM and compare its operating characteristics to those of alternative approaches in two simulation scenarios. In scenario A we consider two exposures and provide a more in depth look at nonlinear exposure-response estimation and interaction detection. In scenario B we include five exposures and evaluate performance for critical window identification and exposure-response estimation with higher-order interactions.

We consider BKMR-DLM with a Gaussian kernel (BKMR-DLM) and with a quadratic kernel (polynomial kernel with d=2; BKMR-DLM-quad). We parameterize the weight functions using a natural spline with four degrees of freedom. For comparison, we also include: 1) BKMR using mean exposure over pregnancy; 2) an additive DLM with natural splines; and 3) an  additive DLNM with penalized splines. 

We used real exposure data taken from one Boston, MA, USA monitor and birth dates simulated uniformly between the years 2007 and 2014. We use data for 5 pollutants: PM$_{2.5}$, nitrogen dioxide (NO$_2$), carbon dioxide (CO), ozone (O$_3$), and sulfur dioxide (SO$_2$) (enumerated $m=1,\dots,5$ in that order). We  simulated births at randomly selected birth dates. For each birth we constructed weekly average exposures for the 37 weeks prior to the simulated birth. This simulation strategy yields a realistic correlation structure both within and across pregnancies.

For both scenarios, we simulated the outcomes using the model 
\begin{equation}
    y_i = h_i + \mathbf{Z}_i^T\boldsymbol{\gamma} + \epsilon_i.
\end{equation}
We simulated  $\mathbf{Z}_i^T=(Z_{i1},\dots,Z_{i5})$ and $\boldsymbol{\gamma}=(\gamma_1,\dots,\gamma_5)^T$ as independent standard normal random variables.  The random error $\epsilon_i$ was simulated as mean zero normal error with standard deviation equal to 3, 7.5, and 15,  which represent approximately a 1:2, 1:5, and 1:10 ratio between the standard deviation of $h$ and $\epsilon$, respectively.  Finally, we considered sample sizes of $n=100$ and $n=500$ and evaluated model performance based on 200 simulated data sets.

We simulated the exposure effect $h_i$ by first creating weighted exposures. The simulated weight functions were a normal density function peaking mid-pregnancy ($w_1$) and a logistic link function identifying a window in the second half of gestation ($w_2$). These represent two biologically plausible ways in which the strength of association may vary over pregnancy.  Both weight functions were truncated to span 37 weeks and scaled to meet the $\int_\mathcal{T}[w(t)]^2dt=1$ constraint (see supplemental Figure~S2 for a visualization). 

In scenario A we included two exposures (PM$_{2.5}$ and NO$_2$) in both the data generating mechanism and the models fit to the simulated dataset. The simulated exposure-response function for scenario A was
\begin{equation}
    h_i = 3/[1+\exp(-2E_1^s)] + 2E_2^s\mathbbm{1}_{(E_2^s>0)} - E_1^sE_2^s \label{eq:simA}
\end{equation}
where $E_1^s$ and $E_2^s$ were scaled and centered versions of the weighted exposures using $w_1$ and $w_2$, respectively. 

In scenario B, we included three exposures and all two-way and three-way interactions between the three active exposures in the data generating mechanism.  The exposure-response function was
\begin{equation}
    h_i = E_{i1}^s+E_{i2}^s+E_{i3}^s + E_{i1}^sE_{i2}^s+E_{i1}^sE_{i3}^s+E_{i2}^sE_{i3}^s + E_{i1}^sE_{i2}^sE_{i3}^s.
\end{equation}
This included a third active exposure, CO. We let $w_3=w_2$ and $E_3^s$ be the scaled centered version of the weighted exposure. We included five exposure in the regression models fit to the simulated datasets, including  O$_3$ and SO$_2$ that had no association with the outcome. 

All models were misspecified. BKMR-DLM was the only model that accounted for exposure-timing, nonlinearity, and interactions. However, the natural spline basis that we employed (4 degrees of freedom) was not sufficiently flexible to model the simulated weight functions and the quadratic kernel was not sufficiently flexible to accurately represent the true exposure-response function in scenario A or the three-way interaction in scenario B.  

\subsection{Evaluation criteria}

We evaluated the ability of each model to estimate the exposure-response function by calculating root mean square error (RMSE) and 95\% interval coverage for $h$. To evaluate interactions in scenario A, we estimate the posterior probability that an interquartile (IQR) change in PM$_{2.5}$ at the $75^\text{th}$ percentile of NO$_2$ is less than an IQR change of PM$_{2.5}$ at the $25^\text{th}$ percentile of NO$_2$. We evaluated each method's ability to estimate critical windows by using the pointwise RMSE for the weight function and 95\% interval coverage for the weight functions. In scenario B, we calculated the proportion of times a critical window was identified for the three active exposure variables and for the two inactive exposure variables. Critical windows were identified as time points where the 95\% interval does not contain zero. We also present the precision of critical window identification, which is the number of correct windows identified divided by total number of windows identified.

To compare the results of DLM and DLNM to the true simulated weight functions, we normalized the estimates to match the constraints imposed on the true weight functions. For DLNM, where the distributed lag function varies smoothly with concentration and windows may only be identified at some concentrations but not at others, we selected the cross-section of the DLNM that shows the association at the mean exposure level for each pollutant and used that cross section to identify windows. 

\subsection{Simulation results}

Tables~\ref{tab:simA} and \ref{tab:simB} compare the performance of the five approaches for Scenarios A and B, respectively. BKMR-DLM-quad performed best in terms of inference on the exposure-response function $h$. For both scenarios, BKMR-DLM-quad had the lowest RMSE on $h$ at the larger sample sizes and larger signal-to-noise levels. BKMR-DLM-quad had coverage for $h$ close to the nominal level, ranging from 0.91 to 0.99 across the settings. In the lower signal-to-noise and smaller sample sizes, BKMR with average exposure had lower RMSE on $h$ but interval coverage for $h$ was well below the nominal level and ranged from 0.41 to 0.93.

BKMR-DLM with the Gaussian kernel also performed well but yielded less efficient estimates of the exposure-response function in the more challenging scenarios. DLM and DLNM  had low interval coverage and had larger RMSE for $h$ than BKMR-DLM with a quadratic kernel.

BKMR-DLM-quad was best able to identify an interaction effect. The mean posterior probability that there was an interaction was 0.963 for scenario A with $n=500$ and high signal-to-noise. The model detected an interaction with high posterior probability in almost all simulated data sets. Performance decreased at smaller sample sizes and lower signal-to-noise ratios. In comparison, BKMR and BKMR-DLM with a Gaussian kernel had mean posterior probability of about 0.5 and low frequency of detecting an interaction. Additive DLM and DLNM are not capable of identifying an interaction.

In scenario A with only two pollutants, BKMR-DLM and BKMR-DLM-quad were best able to estimate the  weight function that characterizes which times have the strongest association with the outcome.   Specifically, BKMR-DLM and BKMR-DLM-quad had the lowest RMSE for the weights (Tables~\ref{tab:simA}).  Supplemental Figures~S2-S5 show the true weight functions and the estimated weight functions for the first 100 simulated data sets in scenario A. Both approaches estimated the general pattern but over-smooth the weight function for the first exposure due to the fact that the spline basis used is not sufficiently flexible to match the peak of the window in the middle of pregnancy. For this scenario,   all of the methods yielded interval coverage for the weights between 0.8 and 0.9. For the two BKMR-DLM approaches, this is due to the lack of flexibility of the spline. For DLM and DLNM, this is due to failure to account for interactions. In scenario B with five exposures, BKMR-DLM and BKMR-DLM-quad had the lowest RMSE for the weight function estimators. 

Table~\ref{tab:simB} compares the critical window identification for scenario B.  DLM and DLNM had the highest frequency of detecting windows in all settings. This came at the cost of a high frequency of detecting false windows. The precision for DLM and DLNM was between 0.6 and 0.7 for all settings. These methods identified a window on an exposure that was not associated with the outcome almost as frequently as they identified a window where there truly was one. In comparison, BKMR-DLM-quad had fairly high probability of detecting correct critical windows in the larger sample size and high signal-to-noise setting while maintaining a very low selection rate for incorrect exposures. This resulting in a very high precision above 0.99.  The true window identification rates with BKMR-DLM with a quadratic and Gaussian kernel decreased for the lower sample size and signal-to-noise settings but the approaches maintain very high precision. Hence, the lower critical window identification rates come with a greater assurance that the windows identified are not spurious.

\begin{table}
\centering
\caption{Simulation results for scenario A with two active exposures. The table shows (from left to right) RMSE for the exposure-response function and 0.95 interval coverage for the exposure-response function, the average pointwise RMSE and pointwise 0.95 coverage for the two weight functions, and the probability that an interaction is detected by comparing the difference in an IQR change in PM$_{2.5}$ at the $75^\text{th}$ and $25^\text{th}$ percentile of NO$_2$. A probability of an interaction near 1 indicates the model consistently finds evidence of an interaction and a probability near 0.5 indicates no evidence of interaction.}
\label{tab:simA}
{\footnotesize
\begin{tabular}{lccccc}\hline
Model	&	RMSE $h$	&	Coverage $h$	&	RMSE  $w(t)$ 	&	Coverage $w(t)$ &	Pr(interact) \\ \hline
\multicolumn{6}{c}{$n=100$,   noise:   $\text{sd}(\epsilon)=3.0$}\\
BKMR & 1.352 & 0.596 & 0.703 &    NA & 0.498  \\ 
BKMR-DLM & 1.241 & 0.975 & 0.603 & 0.887 & 0.502  \\ 
BKMR-DLM-quad & 1.038 & 0.915 & 0.546 & 0.891 & 0.647  \\ 
DLM & 1.205 & 0.857 & 0.896 & 0.865 &    NA  \\ 
DLNM & 1.186 & 0.828 & 0.945 & 0.872 &    NA  \\ 
 \hline
\multicolumn{6}{c}{$n=100$,   noise:   $\text{sd}(\epsilon)=7.5$}\\
BKMR & 1.676 & 0.821 & 0.703 &    NA & 0.496  \\ 
BKMR-DLM & 2.951 & 0.987 & 0.725 & 0.901 & 0.517  \\ 
BKMR-DLM-quad & 1.767 & 0.975 & 0.701 & 0.895 & 0.530  \\ 
DLM & 2.371 & 0.921 & 1.144 & 0.823 &    NA  \\ 
DLNM & 2.224 & 0.903 & 1.127 & 0.845 &    NA  \\  \hline
\multicolumn{6}{c}{$n=100$,   noise:   $\text{sd}(\epsilon)=15.0$}\\
BKMR & 2.283 & 0.927 & 0.703 &    NA & 0.497  \\ 
BKMR-DLM & 5.263 & 0.994 & 0.742 & 0.905 & 0.508  \\ 
BKMR-DLM-quad & 2.942 & 0.994 & 0.746 & 0.895 & 0.516  \\ 
DLM & 4.560 & 0.930 & 1.248 & 0.798 &    NA  \\ 
DLNM & 4.098 & 0.922 & 1.227 & 0.823 &    NA  \\ \hline
\multicolumn{6}{c}{$n=500$,   noise:   $\text{sd}(\epsilon)=3.0$}\\
BKMR & 1.234 & 0.413 & 0.703 &    NA & 0.479  \\ 
BKMR-DLM & 0.639 & 0.928 & 0.469 & 0.822 & 0.506  \\ 
BKMR-DLM-quad & 0.621 & 0.847 & 0.441 & 0.814 & 0.963  \\ 
DLM & 0.881 & 0.681 & 0.712 & 0.883 &    NA  \\ 
DLNM & 0.792 & 0.848 & 0.836 & 0.926 &    NA  \\ 
 \hline
\multicolumn{6}{c}{$n=500$,   noise:   $\text{sd}(\epsilon)=7.5$}\\
BKMR & 1.417 & 0.582 & 0.703 &    NA & 0.478  \\ 
BKMR-DLM & 1.371 & 0.990 & 0.644 & 0.886 & 0.494  \\ 
BKMR-DLM-quad & 1.124 & 0.918 & 0.569 & 0.885 & 0.617  \\ 
DLM & 1.332 & 0.847 & 0.929 & 0.851 &    NA  \\ 
DLNM & 1.394 & 0.881 & 0.964 & 0.894 &    NA  \\ 
 \hline
\multicolumn{6}{c}{$n=500$,   noise:   $\text{sd}(\epsilon)=15.0$}\\
BKMR & 1.662 & 0.762 & 0.703 &    NA & 0.479  \\ 
BKMR-DLM & 2.567 & 0.999 & 0.721 & 0.900 & 0.519  \\ 
BKMR-DLM-quad & 1.677 & 0.970 & 0.689 & 0.893 & 0.538  \\ 
DLM & 2.245 & 0.902 & 1.108 & 0.811 &    NA  \\ 
DLNM & 2.396 & 0.910 & 1.119 & 0.855 &    NA  \\ 
 \hline
\end{tabular}
}
\end{table}

\begin{table}
\centering
\caption{Simulation results for scenario B with three active exposure and two null exposures.  The table shows (from left to right) RMSE for the exposure-response function and 0.95 interval coverage for the exposure-response function, the average pointwise RMSE and pointwise 0.95 coverage for the weight functions for the three active exposures, the frequency of detecting a window in the three active exposures, the frequency of detecting a window in the two exposures not associated with the outcome, and the precision for window detection which is the number of correct windows identified divided by total number of windows identified.}
\label{tab:simB}
{\footnotesize
\Rotatebox{90}{
\begin{tabular}{lccccccc}\hline
Model	& RMSE $h$	&	Coverage $h$ &	RMSE  $w(t)$ 	&	Coverage  $w(t)$ &	Pr(window$|$active) & Pr(window$|$null)  & Precision	\\ \hline
\multicolumn{8}{c}{$n=100$,   noise:   $\text{sd}(\epsilon)=3.0$}\\		
BKMR & 2.087 & 0.758 &    NA &    NA &    NA &    NA &    NA  \\ 
BKMR-DLM & 1.940 & 0.945 & 0.667 & 0.936 & 0.000 & 0.000 &   NaN  \\ 
BKMR-DLM-quad & 1.399 & 0.955 & 0.539 & 0.910 & 0.298 & 0.007 & 0.984  \\ 
DLM & 2.047 & 0.869 & 0.935 & 0.835 & 0.652 & 0.465 & 0.678  \\ 
DLNM & 1.793 & 0.861 & 0.783 & 0.806 & 0.719 & 0.475 & 0.694  \\ 
 \hline
\multicolumn{8}{c}{$n=100$,   noise:   $\text{sd}(\epsilon)=7.5$}\\
BKMR & 2.487 & 0.823 &    NA &    NA &    NA &    NA &    NA  \\ 
BKMR-DLM & 3.392 & 0.978 & 0.682 & 0.937 & 0.000 & 0.000 &   NaN  \\ 
BKMR-DLM-quad & 2.646 & 0.973 & 0.681 & 0.925 & 0.020 & 0.000 & 1.000  \\ 
DLM & 3.814 & 0.915 & 1.092 & 0.860 & 0.385 & 0.324 & 0.641  \\ 
DLNM & 3.496 & 0.894 & 1.012 & 0.814 & 0.430 & 0.332 & 0.661  \\ 
 \hline
\multicolumn{8}{c}{$n=100$,   noise:   $\text{sd}(\epsilon)=15.0$}\\
BKMR & 2.948 & 0.898 &    NA &    NA &    NA &    NA &    NA  \\ 
BKMR-DLM & 5.973 & 0.990 & 0.683 & 0.937 & 0.000 & 0.000 &   NaN  \\ 
BKMR-DLM-quad & 4.480 & 0.989 & 0.727 & 0.926 & 0.000 & 0.000 &   NaN  \\ 
DLM & 7.130 & 0.928 & 1.201 & 0.842 & 0.303 & 0.280 & 0.619  \\ 
DLNM & 6.339 & 0.912 & 1.147 & 0.809 & 0.296 & 0.285 & 0.609  \\ 
 \hline
\multicolumn{8}{c}{$n=500$,   noise:   $\text{sd}(\epsilon)=3.0$}\\
BKMR & 1.354 & 0.819 &    NA &    NA &    NA &    NA &    NA  \\ 
BKMR-DLM & 0.866 & 0.974 & 0.407 & 0.925 & 0.393 & 0.002 & 0.996  \\ 
BKMR-DLM-quad & 0.812 & 0.927 & 0.379 & 0.851 & 0.723 & 0.007 & 0.993  \\ 
DLM & 1.700 & 0.636 & 0.856 & 0.649 & 0.858 & 0.828 & 0.609  \\ 
DLNM & 1.122 & 0.903 & 0.653 & 0.832 & 0.857 & 0.790 & 0.619  \\ 
 \hline
\multicolumn{8}{c}{$n=500$,   noise:   $\text{sd}(\epsilon)=7.5$}\\
BKMR & 2.213 & 0.706 &    NA &    NA &    NA &    NA &    NA  \\ 
BKMR-DLM & 2.261 & 0.983 & 0.661 & 0.936 & 0.000 & 0.000 &   NaN  \\ 
BKMR-DLM-quad & 1.543 & 0.963 & 0.526 & 0.914 & 0.330 & 0.002 & 0.995  \\ 
DLM & 2.300 & 0.823 & 0.923 & 0.837 & 0.693 & 0.485 & 0.682  \\ 
DLNM & 2.147 & 0.906 & 0.813 & 0.878 & 0.622 & 0.435 & 0.682  \\ 
 \hline
\multicolumn{8}{c}{$n=500$,   noise:   $\text{sd}(\epsilon)=15.0$}\\
BKMR & 2.484 & 0.733 &    NA &    NA &    NA &    NA &    NA  \\ 
BKMR-DLM & 3.177 & 0.995 & 0.683 & 0.937 & 0.000 & 0.000 &   NaN  \\ 
BKMR-DLM-quad & 2.493 & 0.978 & 0.641 & 0.927 & 0.030 & 0.000 & 1.000  \\ 
DLM & 3.586 & 0.891 & 1.057 & 0.869 & 0.480 & 0.328 & 0.687  \\ 
DLNM & 3.695 & 0.912 & 0.965 & 0.874 & 0.423 & 0.298 & 0.681  \\ 
 \hline
\end{tabular}
}}
\end{table}

\subsection{Supplementary results and discussion of sensitivity}

We include additional scenarios C and D in the supplemental material. Scenario C is similar to scenario A but has  a smoother true weight function. This weight function can be perfectly modeled with a natural spline with four degrees of freedom. Both BKMR-DLM and BKMR-DLM-quad perform better with a smoother weight function. Using a more flexible basis expansion for the weight functions resulted in higher variance estimates and generally did not perform as well in the simulations. Scenario D is similar to scenario B but includes nonlinear effects and only two-way interaction. Results are similar to scenario B.

\subsection{Summary and key takeaways}

Table~\ref{tab:summary} summarizes methods that can be applied in this setting and the advantages and disadvantages of each approach. BKMR and additive DLM methods fail to account for either exposure timing, nonlinear associations, or interactions. BKMR-DLM is the first method designed to address all three features of the data. 

\begin{table}[]
    \caption{Summary of methods and performance. The first three columns show what data features the method accounts for by design. The last three columns summarize relative performance of the methods and indicates the recommended method based on study objectives of estimating the exposure-response function (ER) and detecting interactions among exposures. These recommendations are based on the simulation study results.}
    \label{tab:summary}
    \begin{tabular}{lcccccc}\hline
          & \multicolumn{3}{c}{Method accounts for by design} 
          & \multicolumn{3}{c}{Method recommended use} \\  \cmidrule(l){2-4}  \cmidrule(l){5-7}
          & \multicolumn{3}{c}{}   & 
          & \multicolumn{2}{c}{Estimating ER} \\  
          & exposure &  inter- & non- & detecting &  wk signal$^*$ &  st signal$^*$ \\ 
         Method &  timing & actions & linearity & interactions &  modest $n$ &  large $n$ \\  
         \hline
         Add. DLM  & $\checkmark$ &              &              &  & & \\
         Add. DLNM & $\checkmark$ &              & $\checkmark$ & & &\\
         BKMR          &              & $\checkmark$ & $\checkmark$ & & $\checkmark$ & \\
         BKMR-DLM      & $\checkmark$ & $\checkmark$ & $\checkmark$ & $\checkmark$ & & $\checkmark$ \\
         \hline
    \end{tabular}
    \begin{flushleft}
    $^*$ denotes weaker signal and stronger signal, respectively.
    
    ER: exposure-response
    \end{flushleft}
\end{table}

BKMR-DLM with the quadratic kernel was best able to estimate the exposure-response relationship in larger sample size and larger signal settings and was the only method that consistently identified interactions among exposures. Even when BKMR-DLM was unable to identify critical windows, accounting for the timing using the flexible weight functions improved exposure-response function estimation. In lower signal settings,  BKMR applied to pregnancy averages was better able to estimate the exposure-response relationship.

Identifying critical windows is challenging. The additive DLM had the most power to identify windows. This came at the cost of identifying an incorrect window about as often as identifying a correct window. BKMR-DLM had lower frequency of identifying critical windows but maintained high precision and almost never identified incorrect windows. Even when BKMR-DLM was unable to identify a critical window, the approach was able to use information on exposure timing to improve exposure-response estimation compared to BKMR with average pregnancy exposures.


\section{Data Analysis}\label{s:da}

We applied BKMR-DLM to analyze BWGAz in the ACCESS cohort. This analysis uses data from the same 109 dyads and four pollutants (nitrate, OC, EC and sulfate) that were analysed using an additive DLM and BKMR based on pregnancy-average exposures in Section~\ref{s:data}. All covariates listed in Section~\ref{s:data} were included in each model. We applied the BKMR-DLM with quadratic kernel. Specifically, the model is
\begin{equation}
Y_i = h(E_{i1},E_{i2},E_{i3},E_{i4}) + \mathbf{Z}_i^T\boldsymbol{\gamma} + \epsilon_i, 
\label{eq:bkmrsingle}
\end{equation}
where each subject-specific exposure above is defined as $E_{im}=\int_\mathcal{T} X_{im}(t)w_m(t)dt$ for $m=1,\dots,4$ ($1=$ nitrate, $2=$ OC, $3=$ EC, $4=$ sulfate). We estimate $h(\cdot)$ with a quadratic kernel as in \eqref{eq:kernelscalarquad} and parameterize the weight functions  using natural splines with four degrees of freedom. Compared to DLM analysis in Section~\ref{sub:dlm}, the exposure-response function allows for interactions and nonlinear associations. Compared to the BKMR model in Section~\ref{sub:kmr}, the input to the exposure-response function is the four vectors of repeated measures of exposures rather than the four scalar pregnancy average exposures. The confounder model $\mathbf{Z}_i^T\boldsymbol{\gamma}$ is the same in all three models. 

\begin{figure}
    \includegraphics[width=\textwidth]{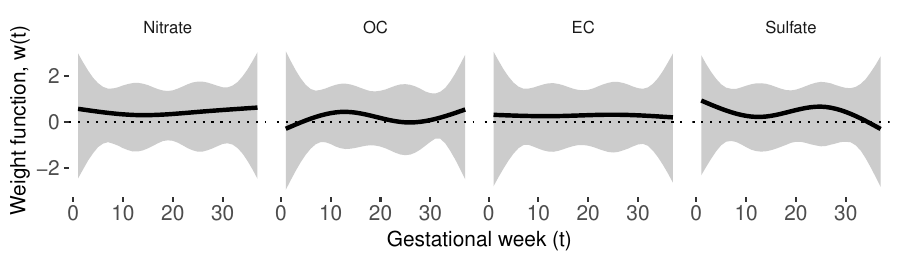}
    \caption{Estimated weight functions from the analysis of BWGAz in ACCESS with BKMR-DLM using a quadratic kernel.  The weight function is constrained and does not reflect the magnitude of the association or the direction of the association. It only reflects the timing of the association.}.
    \label{fig:bkmrdlmwindows}
\end{figure}

Figure~\ref{fig:bkmrdlmwindows} shows the estimated weight functions from BKMR-DLM. No windows were identified. The estimated weight functions do identify periods of time of increased association with the outcome by illustrating periods of time where there is a ``bump'' in the weight function. This increased association is centered at week 12 for OC, which aligns with the critical window for OC identified in the additive DLM analysis in Section~\ref{sub:dlm}. Because the weight function is constrained to meet the identifiability constraints, the sign and magnitude of the weight function do not necessarily correspond to the sign and magnitude of the association. The difference in sign of the weight function compared to that from the additive DLM is, therefore, not indicative of contradicting results.

\begin{figure}
    \centering
    \includegraphics[width=\textwidth]{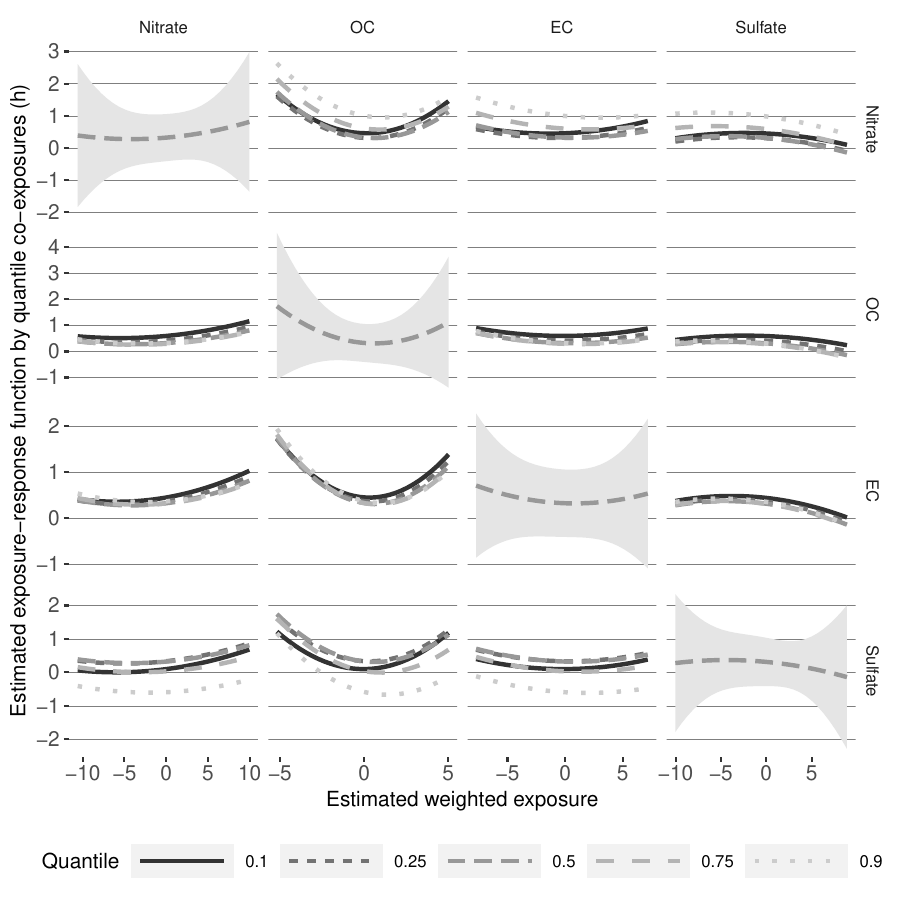}
    \caption{Cross sections of the estimated exposure-response function ($\hat{h}$) from BKMR-DLM with a quadratic kernel. The panels on the diagonal show the main effect, which is the association between a weighted exposure ($x$-axis) and the outcome at the median level of all other weighted exposures. The dashed line represents the posterior mean and the shaded ribbon represents the 0.95 credible interval. The off-diagonals show the exposure-response function at different quantiles of a single co-exposure. For example, the top right panel shows the sulfate exposure-response relationship at different quantiles of nitrate and median levels of OC and EC. A fanning or deviation from parallel lines in the exposure-response relationship represents evidence of an interaction.} 
    \label{fig:bkmrdlmcross}
\end{figure}

Figure~\ref{fig:bkmrdlmcross} shows estimates of the exposure-response function from BKMR-DLM. The diagonal shows $h$ as a function of weighted exposure for one pollutant at the median value of the weighted exposures for the other pollutants. We found some evidence of a negative association between OC and BWGAz. Figure~\ref{fig:bkmrmain} shows a similar estimated association using BKMR with 37-week averaged exposures. The off-diagonals show the posterior mean of $h$ at different quantiles of one co-pollutant and the median of the other two co-pollutants. There was slight evidence that nitrate modifies the OC, EC, and sulfate exposure-response function and that sulfate may modify the OC exposure-response relationship. This interaction was not detected by BKMR using average exposures (Supplemental Figure S1).

\begin{figure}
    \centering
    \includegraphics[width=\textwidth]{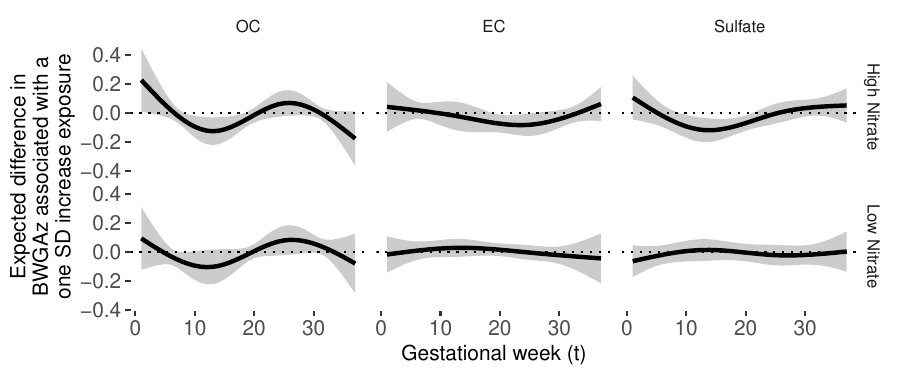}
    \caption{Estimated distributed lag function between exposures and birth weight for gestational age $z$-score in ACCESS using the stratified DLMs. The DLM for each exposure is stratified by mean nitrate level over pregnancy (below and above median nitrate value). The function represents the estimated expected difference in BWGAz per one standard deviation increase in exposure ($y$-axis) as a function of gestational week ($x$-axis).}
    \label{fig:dlmbynitrate}
\end{figure}

Figure~\ref{fig:dlmbynitrate} shows results for univariate DLM analyses of OC, EC and sulfate stratified by nitrate. In this model, we dichotomize Nitrate at the median pregnancy averaged exposure. For each exposure, we included an interaction between the DLM and an indicator of whether pregnancy averaged exposure to nitrate was above or below the median exposure for the cohort. This simpler analysis supports the conclusion that there may be an association between OC, EC, and sulfate at higher levels of nitrate.  Windows of susceptibility associated with OC, EC and sulfate are only observed at higher levels of co-exposure to nitrate. No association between OC, EC and sulfate and BWGAz is observed at lower levels of nitrate.

The data analysis presented is likely under-powered. Using BKMR-DLM, we were able to find some evidence of effect, modification. This identified three (of the 12 possible) two-way interactions to further investigate. By applying the simpler stratified DLM model to these three combinations we were able to identify windows that were not identified with the existing methods. This two-step approach on a small data set can be used to generate hypotheses for further analysis in larger cohorts. For example, the National Institutes of Health (NIH) Environmental Influences on Child Health Outcomes (ECHO) program aims to jointly analyze pooled data from multiple birth cohorts. Such consortium efforts could potentially validate findings found using advanced methods like BKMR-DLM in small cohorts.

Supplemental material Section F presents a BKMR-DLM analysis of the full cohort without limiting the subset of male babies with obese mothers. No critical windows are identified and there is weak evidence of a positive association between nitrate and BWGAz.


\section{Discussion}\label{s:discussion}

In this paper we consider multiple strategies for quantifying the association between time-resolved measures of an environmental mixture and a prospectively assessed  health outcome. In this setting there are three key challenges: accounting for exposure timing, accounting for nonlinear associations, and accounting for interactions.

We proposed BKMR-DLM to estimate the association between time-resolved measures of multiple exposures and an outcome. To our knowledge, this is the first approach that accounts for exposure timing, interactions among exposures, and nonlinear associations--thereby more comprehensively modeling the underlying complexity of the relationships.  The approach uses time-weighted exposures in a kernel machine regression framework.  The weight-functions identify windows of susceptibility during which there is an increased association between exposure and outcome. Such information will be important as developmental processes are both timed and linked to windows of susceptibility; thus exposure timing provides hints to biologic mechanisms underlying health effects.   By using kernel machine regression we allow for nonlinear associations and interactions among the multiple weighted exposures.   

We compared the relative advantages and disadvantages of the proposed approach and simpler approaches through simulation and a case study involving prenatal exposures to multiple air pollutants and birth weight. In a simulation study, we showed that BKMR-DLM with a quadratic kernel function was best able to estimate the exposure-response relationship in most situations. Importantly, even in situations where BKMR-DLM was not powered to unambiguously identify critical windows, including the weighted exposure resulted in improved performance over BKMR using pregnancy average exposure. This was realized through lower RMSE and/or interval coverage closer to the nominal level.

We applied a strategy of combining the use of BKMR-DLM to identify potential interactions and DLM to identify windows of susceptibility to analyze data on prenatal exposure to an air pollution mixture and birth weight in the ACCESS cohort.  In a sample of 109 boys born to obese mothers, we estimated the association between four ambient pollutants and birth weight using BKMR-DLM as a screening method to identify potential interactions between time-resolved exposures. We then performed a stratified DLM analysis to confirm that the findings from the primary BKMR-DLM analysis were not driven solely by modeling assumptions. The fact that the simpler stratified DLM analysis yielded similar results does not diminish the utility of BKMR-DLM, as BKMR-DLM was used to identify which specific pairs of pollutants to further investigate with the stratified DLM. Using a stratified DLM alone would require estimating an unacceptably large number of models when the number of pollutants is large.

There are both limitations and potential future extensions of BKMR-DLM. A first limitation is that the proposed computational approach relies on repeated $n\times n$ matrix inversion, which is computational expensive for large sample sizes. Second, at smaller sample sizes and a lower signal-to-noise ratio the proposed approach lacks power to identify critical windows. Advances in computational approaches for BKMR methods that are currently under development have great promise to scale the approach to larger datasets for which BKMR-DLM will have increased power to detect critical windows. A third limitation is that BKMR-DLM requires complete exposure data and exposure histories of the same length. We therefore limit the exposure data to the first 37 weeks of gestation and include only full term births. This is currently the standard approach in DLM analyses of pre-pregancy exposure data. If exposure is associated with gestational age at birth, including only full term births is conditioning on a mediator and could cause bias. Future research should pursue methods that can handle exposure data of different lengths to accommodate varying gestational ages, including preterm births. Another area for future research is the development of alternative constraints on the weight functions. In this paper we allow each weight function to take positive and negative values. This allows for an effect to be protective at some times and detrimental in other time periods, which has been shown to be the case for some exposures \citep{Bauer2017,ClausHenn2018,Liu2018a}. In cases in which the substantive question involves exposures that are likely to have effects that do not change direction over the time windows under study, an alternative approach would be to implement a  positivity constraint on the weights, $w(t)>0$ $\forall t\in\mathcal{T}$. This would result in the effect being in the same direction at all time points and may increase the model's power to detect windows if the assumption is valid. Finally, data on mixtures observed at repeated time points results in exposure measures that are highly correlated. In cases where the correlation among exposures is extremely high, BKMR-DLM and most other approaches will have limited ability to differentiate the effects of the individual exposures.

Both the estimation of health effects associated with multi-pollutant mixtures and the identification of windows of susceptibility  are important areas of environmental health research. BKMR-DLM integrates these two important areas of research by simultaneously estimating windows of susceptibility and multi-pollutant exposure-response functions.


\section*{Acknowledgements}

This work was supported in part by NIH grants R01ES028811, R01ES013744, P30ES000002, P30ES023515, and UH3OD023337 and US EPA grant RD-83587201. Its contents are solely the responsibility of the grantee and do not necessarily represent the official views of the US EPA.  Further, US EPA does not endorse the purchase of any commercial products or services mentioned in the publication. The ACCESS cohort has been supported by  NIH grants R01ES010932, U01HL072494, and R01HL080674. This work utilized the RMACC Summit supercomputer, which is supported by the NSF (awards ACI-1532235 and ACI-1532236), the University of Colorado Boulder and Colorado State University. 

\section*{Supplement}
Supplemental material available upon request. The supplement contains: A) additional figures for the preliminary analyses in Section~\ref{s:data}. B) Additional details on the method. C) additional simulation study scenarios and results. D) Additional results for the data analysis. F) Data analysis with full cohort.

Software for BKMR-DLM is available with the {\tt bkmrdlm} function in the {\tt R} packages {\tt regimes} available at \url{http://anderwilson.github.io/regimes/}.

\bibliography{TechReport}

\end{document}